      [1999/12/01 v1.4c Il Nuovo Cimento]
\documentclass{cimento}
\usepackage{graphicx} 
\title{STAR results and perspectives on transverse spin asymmetries}
\author{N.~Poljak for the STAR collaboration\from{ins:x}}
\instlist{\inst{ins:x}University of Zagreb, Croatia}
\PACSes{\PACSit{13.88.+e; 13.85.Ni; 12.38.Qk; 14.20.Dh}{}}
\begin{document}

\maketitle

\begin{abstract}
The STAR collaboration reported precision measurements on the transverse single spin asymmetries for the production of forward $\pi^0$ mesons from polarized proton collisions at $\sqrt{s}=$ 200 GeV. To disentangle contributions to measured forward asymmetries one has to look beyond inclusive $\pi^0$ production to the production of forward jets or direct photons. In 2006, STAR with the Forward Pion Detector++ (FPD++) in place, collected 6.8 pb$^{-1}$ of forward data with an average polarization of 60\%. FPD++ had sufficient acceptance for ``jet-like'' objects, which are clustered responses of an electromagnetic calorimeter primarily sensitive to incident photons, electrons and positrons. For these objects, the angle of the outgoing leading $\pi^0$ with respect to the fragmenting parton was reconstructed, thus enabling us to disentangle the contributions to the forward $\pi^0$ asymmetries. The simulated data set shows that on average there are approximately 2.5 fragmenting mesons per one ``jet-like'' object, making them reasonably ``jetty''. Preliminary results provide no evidence of measured contributions to the asymmetry from jet fragmentation, implying the Sivers distribution functions play a substantial role in producing the large inclusive forward $\pi^0$ asymmetries. A similar effort was made in the mid-rapidity $(|\eta|<1)$ region of the STAR detector, where 2.2 pb$^{-1}$ of data was collected. We present progress made by making measurements of the azimuthal asymmetry of leading charged pions in jets produced by transversely polarized proton collisions. 

\end{abstract}

\section{Introduction}
For a particle produced in a collision of transversely polarized protons, the analyzing power $A_N$ is equal to the difference of spin-up and spin-down cross sections divided by their sum. $A_N$ is just one example of a transverse single spin asymmetry, which is expected to be near zero in a leading-twist collinear perturbative QCD description of particle production \cite{ref:Kane78}. In the high-rapidity region, the measured cross sections for production of neutral pions ($\pi^0$) produced with large Feynman-x (2$p_L /\sqrt{s}$) and moderate $p_T$ in pp collisions are found to be in agreement with next-to-leading order pQCD calculations at $\sqrt{s}=200\,$GeV and are included in global fits on fragmentation functions (FFs) \cite{ref:Adams06, ref:deFlorian07}. ``Jet-like" structures were found in two-particle correlations involving a forward pion, as expected from pQCD. Precision measurements of the forward $\pi^0$ asymmetry as a function of $x_F$ and $p_T$ were reported, showing large $A_N$ at large $x_F$ \cite{ref:Abelev08}. The measured $x_F$ dependence is in qualitative agreement with the Sivers effect \cite{ref:Sivers90, ref:Sivers91} expectations. Theoretical understanding of the forward $\pi^0$ production data continues to evolve \cite{ref:Kang11}.

Several classes of models try to explain the observed asymmetries. Simulations indicate that the kinematics of observed events come from processes that reside firmly in the Transverse Moment Distribution (TMD) regime. Hence, we will only discuss two possible contributions to the observed asymmetries within the TMD framework. The Sivers effect manifests itself as an asymmetry in the forward jet or gamma production while the Collins effect \cite{ref:Collins93} manifests itself as an asymmetry in the forward jet fragmentation. Both effects introduce a transverse scale $k_T$ to produce the observed asymmetries. The Collins mechanism introduces $k_T$ in the FFs, making it sensitive to correlations of the hadron transverse momenta and parton transversity. To distinguish between the mechanisms, one has to look beyond inclusive $\pi$ events to direct gamma or ``jet-like'' objects. Since the Collins effect is a spin-dependent azimuthal modulation of hadrons around the thrust axis of an outgoing quark, integrating over the full azimuthal angle would cancel it, leaving only the Sivers effect responsible for any measured asymmetry.

The forward $(<\eta>\approx 3.3)$ electromagnetic detector present during the RHIC running in the year 2006 was specifically designed to have sufficient acceptance for ``jet-like'' objects. A ``jet-like" object is a clustered response of an electromagnetic calorimeter primarily sensitive to incident photons, electrons and positrons.  Theoretical predictions for the estimated $\pi^0$ Collins contribution to the asymmetry for the observed process $(pp \rightarrow \pi^0 + \textrm{``jet-like''} + X)$ show it to be negligible \cite{ref:dAlesio11}. Due to isospin symmetry the $\pi^0$ FFs are half the sum of those for charged pions (which are similar in size and of opposite sign), making the $\pi^0$ Collins FF very small. The data obtained were used to separate the Sivers/Collins contributions to the asymmetries observed for the $\pi^0$ events. 

A similar exploratory analysis was done at the mid-rapidity region $(|\eta|<1)$ of the STAR detector, which is well suited for full jet reconstruction. Observation of the azimuthal distribution of charged pions inside hadronic jets would in a similar way allows one to find the Collins moment of any measured asymmetries.

\section{Experimental setup}

The forward STAR calorimeters prior to 2006 measured the inclusive $\pi^0$ cross section as well as the single beam spin asymmetry for their inclusive production. In 2006 the detectors placed on the West STAR platform were upgraded to a detector called the Forward Pion Detector ++ (FPD++) (Fig.1). A detailed discussion of the detector setup and calibration can be found in \cite{ref:Poljak11}.

The mid-rapidity $(0 <\phi< 2\pi, -1 < \eta < 2)$ section of the STAR detector (Fig.1) consists of a large solid angle detector  well-suited for full jet reconstruction \cite{ref:Ackerman03}. The Time-Projection Chamber (TPC) provides charged particle identification and momentum measurement \cite{ref:Anderson03}. The Barrel and Endcap Electromagnetic Calorimeters (BEMC, EEMC) detect neutral particles and serve for triggering \cite{ref:Beddo03, ref:Allgower03}. TPC and EMC acceptances allow charged particle and jet detection over a pseudorapidity $|\eta| < 1$ in the mid-rapidity analysis. The Beam-Beam Counter ($3.3 < |\eta| < 5$) measures polarization direction, luminosity and provides a minimum bias trigger.

To aid the understanding of the unpolarized results, full PYTHIA/GEANT \cite{ref:Pythia01} simulations have been made with adequate statistics. In forward analysis, we used PYTHIA 6.222, which predates tunings related to ``underlying event'' for midrapidity Tevatron data, since these tunings impact forward production at RHIC energies. The newer versions of PYTHIA, used for mid-rapidity analysis, reduce agreement between data and simulation in the forward region. Energy deposition computed by GEANT for PYTHIA generated events was digitized and was run through the same algorithms as the data.

\begin{figure}[h!t!]
\begin{minipage}{14pc}
\includegraphics[width=14pc]{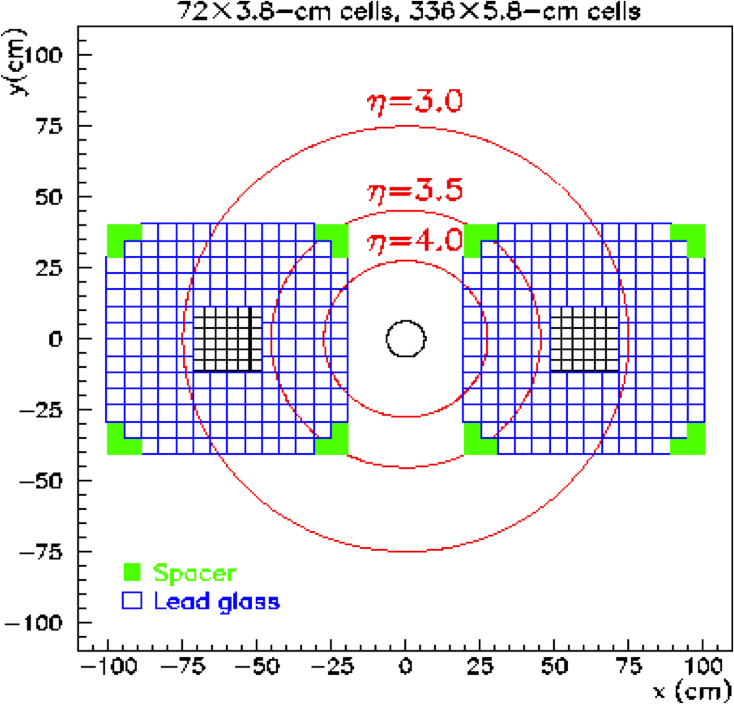}
\end{minipage}\hspace{3pc}
\begin{minipage}{14pc}
\includegraphics[width=14pc]{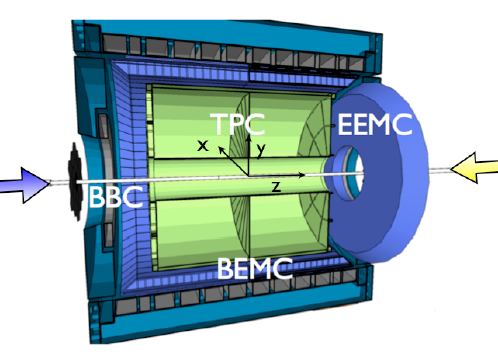}
\end{minipage} 
\caption{\label{Scheme} Left panel: The schematic view of the front face of the FPD++ forward electromagnetic calorimeter.  Right panel: The schematic view of the mid-rapidity part of the STAR detector, showing the STAR coordinate system and four of the essential detector components. The FPD++ is located to the right of the EEMC, outside of the mid-rapidity part.} 
\end{figure}

\section{Results}

In the analysis of the forward data, we formed ``jet-like'' clusters in an event by considering energy depositions $>$ 0.4 GeV in all FPD++ cells. A cluster consists of N cells, where N is the maximal subset of cells found to be within a cone of radius 0.5 in $\eta - \phi$ space. Assuming that the energy deposition is from photons originating from the collision vertex, the $x_F$ and $p_T$ for the cluster are given by the vector sum of momenta from each cell. Clusters with at least 10 cells having cluster $p_T >$ 1.5 GeV/c and $x_F >$ 0.23 are required in the analysis. A requirement that the ``jet-like'' cluster centroid is within the calorimeter volume by at least two large cell widths is also imposed. 

The energy deposition profiles and the invariant mass distributions of the ``jet-like'' objects were found both in data and simulations. Throughout the $x_F$ range of the event samples the properties of the ``jet-like'' objects in the data and the simulations compared well. A subset of data containing both a reconstructed $\pi^0$ and a ``jet-like" object in a single event was extracted. Events from this subset were characterized with the help of PYTHIA, showing that on average the $\pi^0$ carries $\approx 90$\% of the energy of the ``jet-like'' object (Fig.2). Further, the reconstructed ``jet-like'' thrust axis (reconstructed from the energy deposition in the calorimeter) agrees well with the direction of either a hard-scattered parton or a radiated parton. On average, there are 2.5 fragmenting mesons per one ``jet-like'' object, making the objects reasonably jetty.  

The component of the $\pi^0$ momentum perpendicular to the ``jet-like'' object axis ($k_T$) was found to be in the domain of TMD fragmentation (Fig.2). Systematic studies of the ``jet-like'' object model ensured that no special point in the model parameter space was selected. By changing the model parameters by 10\% both on data and simulations, we found that the changes in the unpolarized results are small and smooth. The data and simulations follow the same trends when changing a single parameter.

\begin{figure}[h!t!]
\begin{minipage}{14pc}
\hskip 3mm
\includegraphics[width=14pc]{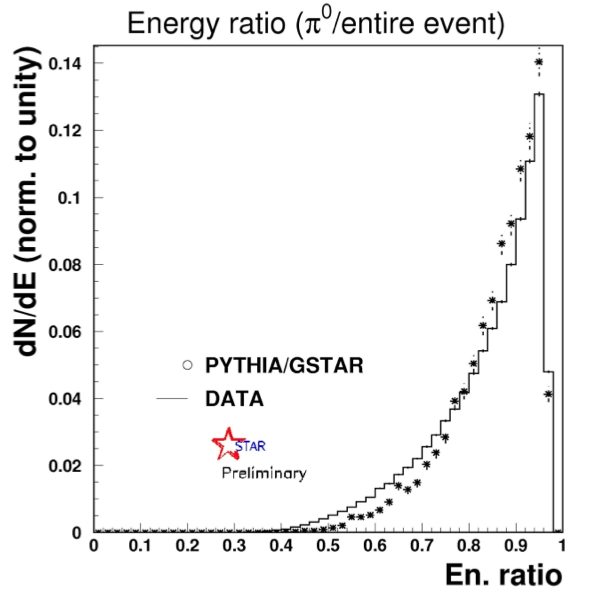}
\end{minipage}\hspace{3mm}
\begin{minipage}{14pc}
\hskip 5mm
\includegraphics[width=14pc]{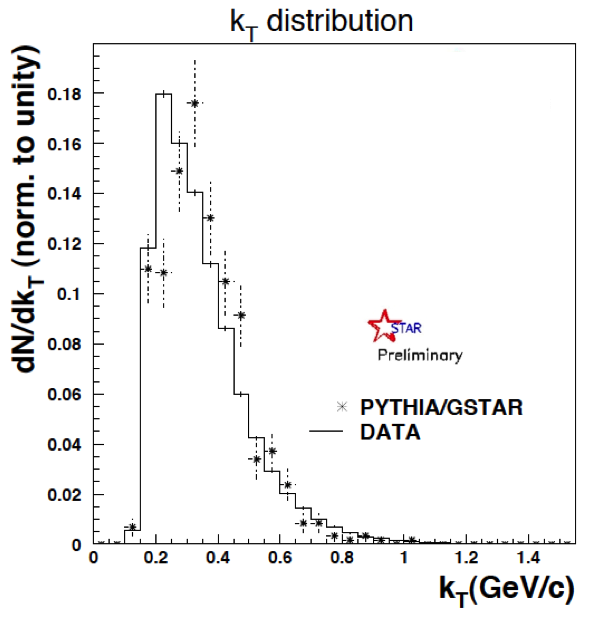}
\end{minipage} 
\caption{Left panel: The fraction of the event energy carried by the leading $\pi^0$ in data and simulations. Right panel: The component of the $\pi^0$ momentum perpendicular to the ``jet-like'' object axis ($k_T$). In both panels the error bars are statistical.}
\end{figure}

To separate the Collins/Sivers contributions to the measured asymmetries, we defined the angle $\gamma$ as the azimuthal angle of the $\pi^0$ with respect to the reaction plane (Fig.3). $\gamma$ is defined mirror symmetrically (clockwise for the left module, counter-clockwise for the right module), so that for both detector modules $\gamma \approx 0$ corresponds to $\pi^0$ having larger $p_T$ than the ``jet-like'' object. For a given bin in $\gamma$ defined this way, an event reconstructed in the left module from an up-polarized proton measures the same fragmentation as an event reconstructed in the right module from a down-polarized proton. 

The unpolarized $\gamma$ angle distribution was found both in data and simulations and shows reasonable agreement \cite{ref:Poljak11}. The peaking near $\gamma =0$ in the shape of the distribution is a detector acceptance effect. In most cases both the $\pi^0$ and the thrust axis of the ``jet-like'' object are reconstructed in the small cells. The $\chi^2$ for fitting a constant to the spin-averaged $\gamma$ distribution systematically decreases in an analysis that restricts the acceptance to have the thrust axis of the ``jet-like'' object increasingly centered on a calorimeter module. Consequently, the shape of the spin-averaged $\gamma$ distribution is due to finite detector acceptance. The $p_T$ dependence of the ``jet-like'' production cross section strongly favors low $p_T$ events. Coupled with detector acceptance, this enhances the number of events with $\gamma \approx 0$. 

The dependence of the asymmetry on the $\gamma$ angle was found. The data was separated in 4 equally sized bins in $\cos(\gamma)$. For each of the bins, the asymmetry was calculated by means of a cross-ratio formula:

\begin{eqnletter}
 \textrm{A}_{\textrm{N}}f(\gamma)=\frac{\sqrt{N_L^{\uparrow}N_R^{\downarrow}}-\sqrt{N_L^{\downarrow}N_R^{\uparrow}}}{\sqrt{N_L^{\uparrow}N_R^{\downarrow}}+\sqrt{N_L^{\downarrow}N_R^{\uparrow}}}
\end{eqnletter}

Here, $N$ stands for the number of measured events in a given bin, for a given module ($L$ or $R$) and a given polarization direction ($\uparrow$ or $\downarrow$). By forming a geometric mean in each $\gamma$ bin, the detector effects are minimized. On the plot of $A_N$ vs. $\cos(\gamma)$ (Fig. 3) the Collins contribution is proportional to the slope parameter, since then the fragmentation is $\gamma$ angle-dependent. If the slope is found to be consistent with zero, while the asymmetry is larger than zero, one has isolated the Sivers effect, since fragmentation is not involved, and one is left with just the initial state $k_T$. On the figure, the lines on the points represent statistical errors and the black area the systematic errors. The $x_F>0$ and the $x_F<0$ points have been slightly horizontally shifted to improve visibility. The asymmetry for negative $x_F$ values is consistent with zero, while the positive $x_F$ values show a positive asymmetry with a significance of 1$\sigma$, but no dependence on $\cos(\gamma)$. Hence, the Collins effect is not found to be present in the production of forward neutral pions. 

\begin{figure}[h!t!]
\begin{minipage}{16pc}
\includegraphics[width=16pc]{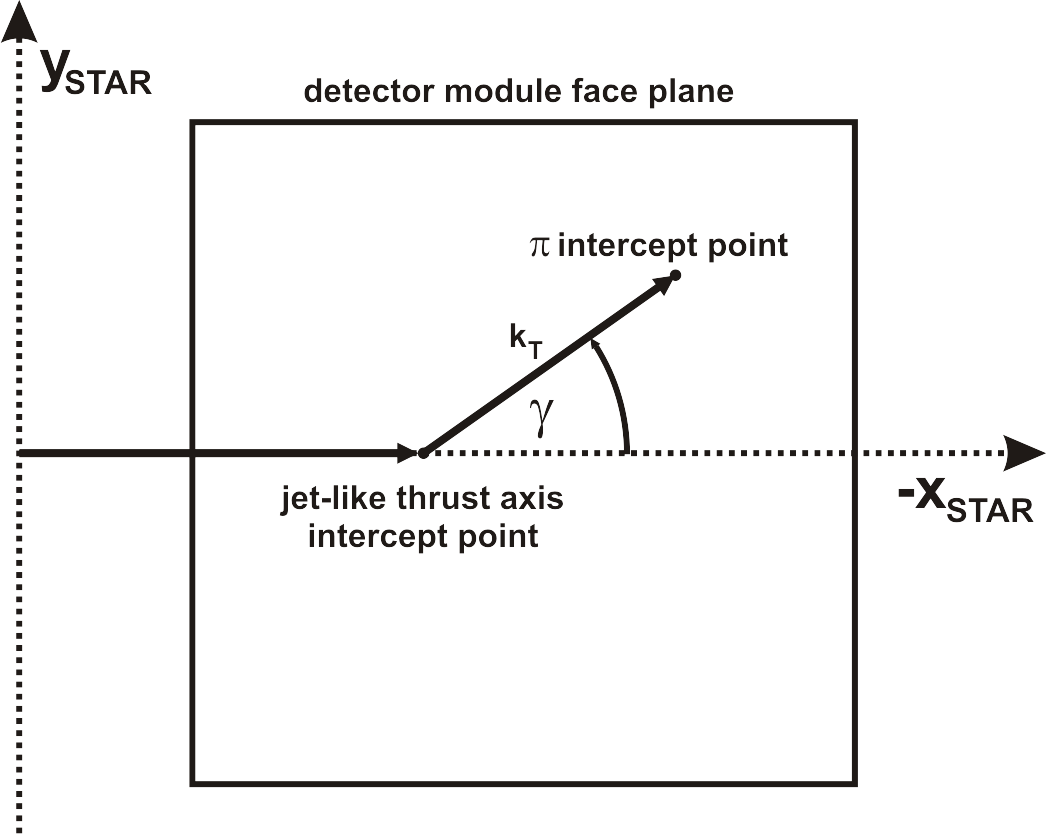}
\end{minipage}\hspace{1mm}
\begin{minipage}{16pc}
\includegraphics[width=16pc]{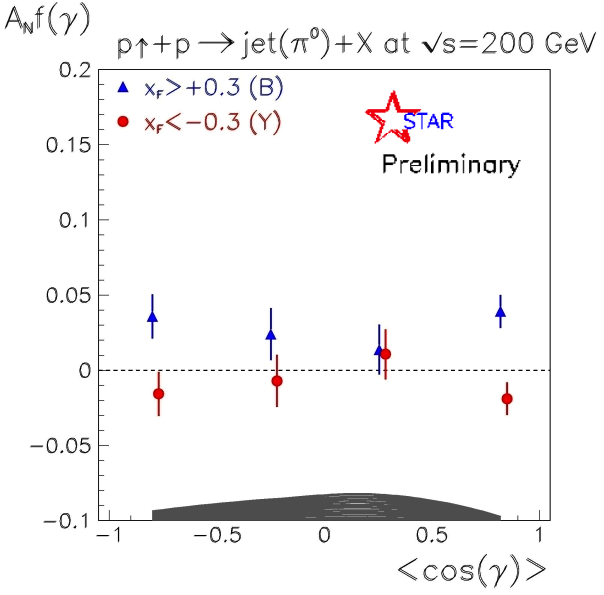}
\end{minipage} 
\caption{Left panel: Schematic view of the $\gamma$ angle in the right detector module. Right panel: The dependence of the asymmetry on the $\gamma$ angle. The lines on the points represent statistical errors and the gray area the systematic errors.}
\end{figure}

To reconstruct jets from the mid-rapidity data, the Midpoint Cone Algorithm was used. A TPC track or EMC tower with an energy of 0.5 GeV or become the seed for a jet with cone radius of 0.7 in the $\eta-\phi$ space. Event triggers required a minimum ADC sum threshold for one of 12 separate ``jet patches''. A cut of $p_T >10$ GeV is imposed to optimize access to quark-scattering events. Charged pions were identified in the TPC. Cuts on the number of track points and the distance of closest approach to their common vertex define an acceptable charge track. A cut of fixed width in $n_{\sigma}$, the Gaussian distribution of $\log(\textrm{d}E/\textrm{d}x)$ about a momentum-dependent centroid, is used to separate charged pions from other charged particles. In the analysis, only charged pions that are leading particles are kept. Only jet triggers in the forward half of the detector with respect to either RHIC polarized beam are analyzed. The Collins moment of the asymmetry (Fig. 4) is measured with respect to either $k_T$ (denoted by $j_T$ in this analysis) or $z (=p_{\pi}/p_{\textrm{jet}})$. The measurements have large uncertainties and we expect significant improvements in the near future. Further details of the analysis can be found in \cite{ref:Fersch2011}.

\begin{figure}[h!t!]
\hskip 3mm
\includegraphics[width=28pc]{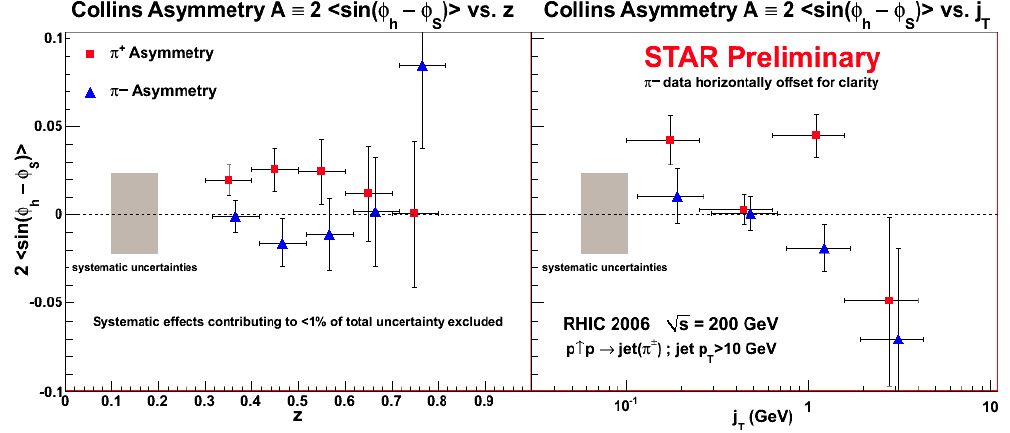}
\caption{The dependence of the measured mid-rapidity Collins asymmetry on $z$ and $j_T$. The lines on the points represent statistical errors and the gray area the systematic errors. Details of the Collins moment calculation (on the vertical axis) can be found in \cite{ref:Fersch2011}.}
\end{figure}

\section{Conclusion and Outlook}

The FPD++ was used to analyze forward ``jet-like'' objects. The data show good agreement with the simulated sample of events. The ``jet-like'' objects were characterized and were shown to be ``jetty''. For the event sample containing both a reconstructed $\pi^0$ and a ``jet-like'' object, the $\gamma$ angle and $k_T$ were found. The $k_T$ was shown to be in the domain of TMD fragmentation. Preliminary results from the correlation analysis of pions and ``jet-like'' objects provide no evidence for a Collins signature. Analysis of mid-rapidity jets has large uncertainties, expected to be improved in the near future. 

For the forward rapidity data, we aim to open up the $\pi^0$ acceptance from small cells to large and small cells in both calorimeter modules, thereby increasing the statistics of the sample and further addressing the $\gamma$ distribution shape. We aim to improve the mid-rapidity uncertainties with inclusion of additional simulation statistics and new analysis methods.

\end{document}